\documentclass[%
 reprint,
 amsmath,amssymb,
 aps, twocolumn
]{revtex4-2}

\usepackage{graphicx}
\usepackage{bm}
\usepackage{xcolor}
\begin{document}

\preprint{APS/123-QED}

\title{Origin of two distinct stress relaxation regimes in shear jammed dense suspensions}

\author{Sachidananda Barik}
\author{Sayantan Majumdar}
 \email{smajumdar@rri.res.in}
\affiliation{Soft Condensed Matter Group, Raman Research Institute, Bangalore 560080\\
}%




\date{\today}

\begin{abstract}
Many dense particulate suspensions show a stress induced transformation from a liquid-like state to a solid-like shear jammed (SJ) state. However, the underlying particle-scale dynamics leading to such striking, reversible transition of the bulk remains unknown. Here, we study transient stress relaxation behaviour of SJ states formed by a well-characterized dense suspension under a step strain perturbation. We observe a strongly non-exponential relaxation that develops a sharp discontinuous stress drop at short time for high enough peak-stress values. High resolution boundary imaging and normal stress measurements confirm that such stress discontinuity originates from the localized plastic events, whereas, system spanning dilation controls the slower relaxation process. We also find an intriguing correlation between the nature of transient relaxation and the steady state shear jamming phase diagram obtained from the Wyart-Cates Model.
\end{abstract}
\maketitle
Stress induced enhancement of viscosity in dense particulate suspensions, commonly known as shear-thickening \cite{barnes1989shear, hoffman1972discontinuous, wagner2009shear, brown2014shear}, has attracted significant recent interests both from the fundamental points of view and materials design \cite{brown2010generality, wyart2014discontinuous, seto2013discontinuous, mari2014shear, guy2015towards, lee2003ballistic, majumdar2013optimal}. For high enough particle volume fraction and applied stress, many of these systems show a remarkable transition to a shear jammed (SJ) state showing stress-activated solid-like yield stress \cite{peters2016direct, majumdar2017dynamic, han2016high}. Although, the phenomenon of shear-thickening is well-known since past few decades, the difference between SJ state and strong/discontinuous shear-thickened (DST) state has been demonstrated only recently and remains a topic of intense research \cite{peters2016direct, singh2019yielding, guy2018constraint, james2018interparticle, dhar2019signature}. SJ results from the stress induced constraints on sliding and rolling degrees of freedom of particles \cite{guy2018constraint, singh2019yielding, singh2020shear}. Such constraints can originate from frictional, hydrodynamic and other short-range inter-particle interactions \cite{james2018interparticle, singh2019yielding, jamali2019alternative}. For steady state flow of frictional systems, Wyart-Cates (W-C) Model provides microscopic insight into the stress ($\sigma$) induced increase in viscosity in terms of a single order parameter $f = f(\sigma)$: the fraction of frictional contacts in the system \cite{wyart2014discontinuous}. A recent continuum model \cite{baumgarten2019general} that treats the parameter `$f$' as a spatially varying field with specific time-evolution can quantitatively describe a range of striking flow behaviours in these systems. Once the applied perturbation is removed, SJ state quickly relaxes back to the unperturbed fluid-like state. Such fast reversibility coupled with high stress bearing ability of SJ systems \cite{brown2014shear}, remain at the heart of many potential applications. 

The nature of relaxation in dense suspensions close to jamming is complex and is sparsely explored in the context of systems showing DST and SJ: Above a critical deformation rate, dense suspensions of PMMA nano-paricles show two distinct relaxation processes \cite{d1993scattering}, in dense cornstarch suspensions the relaxation behaviour deviates significantly from a generalized Newtonian model \cite{maharjan2017giant} and many aspects of the relaxation can be captured by a continuum theoretical model \cite{baumgarten2020modeling, baumgarten2019general}. In dense PSt-EA nano-particle suspensions, a multi-element viscoelastic model consistent with the expected force chain structure fits the relaxation behaviour well \cite{cao2018stress}. Two-step relaxation have also been observed in glassy and static-jammed materials \cite{mohan2015build, chen2020microscopic}. However, extremely slow nature of relaxation and the existence of residual stresses in these systems highlight the widely different underlying microscopic dynamics as compared to the dense suspensions showing DST and SJ.

Despite these detailed studies, the role of microscopic particle scale dynamics in controlling the bulk relaxation of SJ state is not understood. Importantly, due to its solid-like nature, a sustained steady state flow is not possible in SJ systems \cite{dhar2019signature}. Thus, in earlier studies the mechanical state of the SJ sample just prior to relaxation remains poorly characterized. Recent experiments suggest that the mechanical properties of SJ states can be probed reliably under transient perturbations \cite{waitukaitis2012impact, peters2016direct, majumdar2017dynamic, james2018interparticle}. Numerical simulations probing transient relaxation in over-damped, athermal dense suspensions of frictionless spheres found a power-law cut-off by an exponential relaxation close to the isotropic jamming point ($\phi_0$) with the relaxation time diverging at $\phi = \phi_0$ \cite{hatano2009growing, ikeda2020universal}. Although, the relevance of such findings for SJ systems remains unclear due to the difference in stress dependence of the constraints in frictionless and frictional systems \cite{guy2018constraint}.    

In this Letter, we address these issues by studying the transient stress relaxation behaviour of SJ states in dense suspensions of colloidal polystyrene particles (PS) dispersed in polyethylene glycol (PEG) using shear rheology in conjugation with high resolution boundary imaging. We observe a power-law cut-off by a stretched-exponential relaxation  for moderate stress values, however, a sharp discontinuous relaxation after the power-law regime is observed for large stresses. We directly correlate, for the first time, plasticity and dilation in the system with the time scales associated with the bulk relaxation dynamics. We also uncover an interesting connection between the transient relaxation phenomena and the steady state SJ (SSSJ) phase diagram obtained using the W-C model. 

The shear thickening dense suspensions are prepared by dispersing mono-disperse PS (diameter = 2.65$ \pm 0.13\,\mu$m) in PEG \cite{dhar2019signature} for a range of volume fractions. See Supplementary Information (S.I.) \cite{Suppl} for details, which also includes Refs. \cite{wang2013synthesis, singh2018constitutive, ref, utter2004self, besseling2007three, olsson2010diffusion, eisenmann2010shear}. 
\begin{figure}
\includegraphics[scale=0.5]{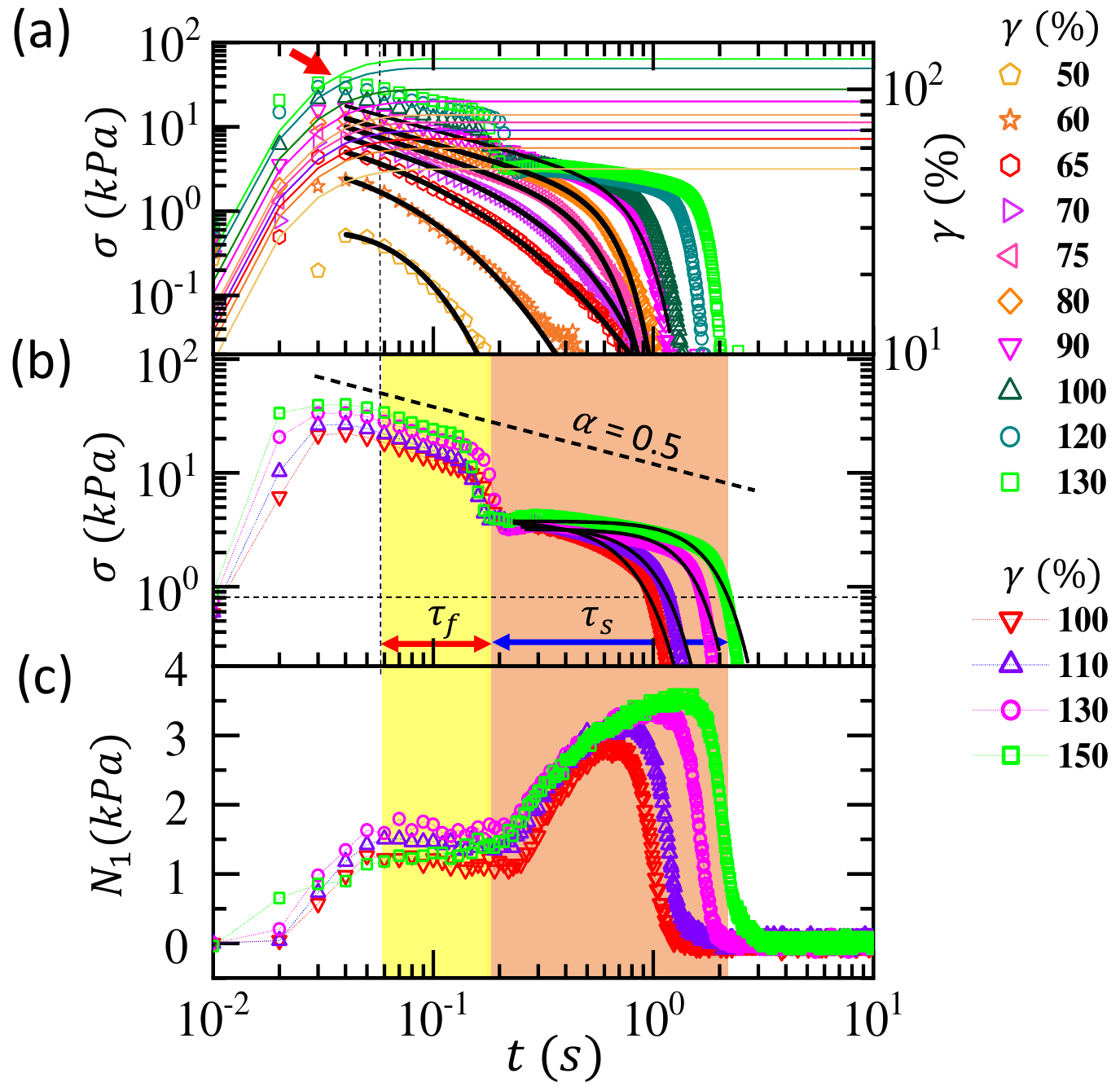}
\caption{\label{F1}(a) Shear stress $\sigma$ as a function of time $t$ (symbols) under different applied step strains $\gamma$ (thin lines). Solid lines are the fits to the power-law cut-off by a stretched exponential function (main text). The arrow indicates peak stress ($\sigma_p$) for $\gamma$ = 130\%. (b) Plots of $\sigma$ vs $t$ for $\gamma \geq 100$\%  showing a self similarity. Power-law decay with slope 0.5 is also indicated. Solid-lines show typical stretched exponential fits after the discontinuous stress drop. The fast $\tau_f$ (yellow shade) and slow $\tau_s$ (brown shade) relaxation time scales are marked. (c) Evolution of first normal stress difference $N_1$ corresponding to the data shown in panel (b). In all cases $\phi = 58\%$ and solvent viscosity $\eta_l$ = 80 $mPa.s$.}
\end{figure}
        As a transient perturbation, a step shear strain of a certain magnitude ($\gamma$) is applied to the sample for 25 s. We record the resulting stress response. Due to instrumental limitation the applied strain reaches the set value after a time $t_0 \approx$ 0.06 s as shown in Fig.1(a). Shear stress in the system quickly reaches a maximum ($\sigma_{p}$) and then starts to relax. We find that the nature of stress relaxation is determined by the magnitude of $\sigma_{p}$ that depends on both $\phi$ and $\gamma$. We observe that for $57\% \leq \phi \leq 61\%$ ($\phi$ range corresponds to SSSJ, see Fig.S1), the stress relaxation is given by $\sigma(t) \sim t^{-\alpha}\, e^{-(t/\tau)^{\beta}}$ for values of $\sigma_p$ up to $\sim$ 16 kPa. Remarkably, for $\sigma_p >$ 16 kPa, we find a discontinuous stress drop soon after the power-law decay regime by almost an order of magnitude (Fig.1(a) and 1(b)). After this, the stretched exponential function still captures the long-time relaxation fairly well. Similar trends are also observed for other $\phi$ values in SSSJ regime but, the magnitude of the discontinuous stress drop increases for larger $\phi$ values (Fig.S2(a) and S2(b)). In all cases, the magnitude of power law exponent ($\alpha$) decreases with increasing $\sigma_p$ and finally saturates at $\alpha \approx 0.5$ except for $\phi$ = 57\%, where we get a higher saturation value (Fig.S3(a)). Just below SJ ($\phi <$ 57\%), the power-law regime disappears and a stretched exponential relaxation is observed for a wide range of $\sigma_p$ (Fig.S3(b)). Interestingly, we observe similar  relaxation dynamics for a variety of dense suspensions showing SJ (Fig.S3(c)) indicating an universal behaviour. For quantification, we define two time-scales for such discontinuous stress relaxation: a fast time scale $\tau_f$ (after $t_0$) at which the discontinuous stress drop takes place and a slower one $\tau_s$ indicating the time (after $\tau_f$) for the stress to drop to $\sigma_d = 0.05\,\Gamma / a$ \cite{brown2012role} ($\Gamma$: surface tension of solvent-air interface, $a$: particle diameter). Notably, most of the stress in the system relaxes within $t = \tau_f$. In Fig.1(c) we show the variation of first normal stress difference $N_1 = \frac{2\,F_N}{\pi\,r^2}$ (for cone-plate geometry where, $F_N$ is the normal force on the plate/cone) corresponding to the stress relaxation data shown in Fig.1(b). In all cases we find that $N_1$ shows a clear positive peak at a longer time near $\sigma = \sigma_d$, with the instantaneous shear stress $\sigma(t) \approx N_1$ near the peak. On the other hand, the behaviour of $N_1$ remains arbitrary (Fig.1c and also Fig.S2(a) and S2(b)) for $t < \tau_f$ and $\sigma(t) >> N_1$ in this regime. Such interesting decoupling of the shear and the normal stress response is not observed for SSSJ (Fig.S1) \cite{royer2016rheological}. For smaller $\sigma_p$ values (smooth relaxation regime) we find that the magnitude of the peak in $N_1$  monotonically decreases with decreasing $\sigma_p$. Also, the peak gradually shifts towards the smaller time scales and completely disappears for sufficiently small values of $\sigma_p$ (Fig.S4).  
\begin{figure*}
\includegraphics[scale=.46]{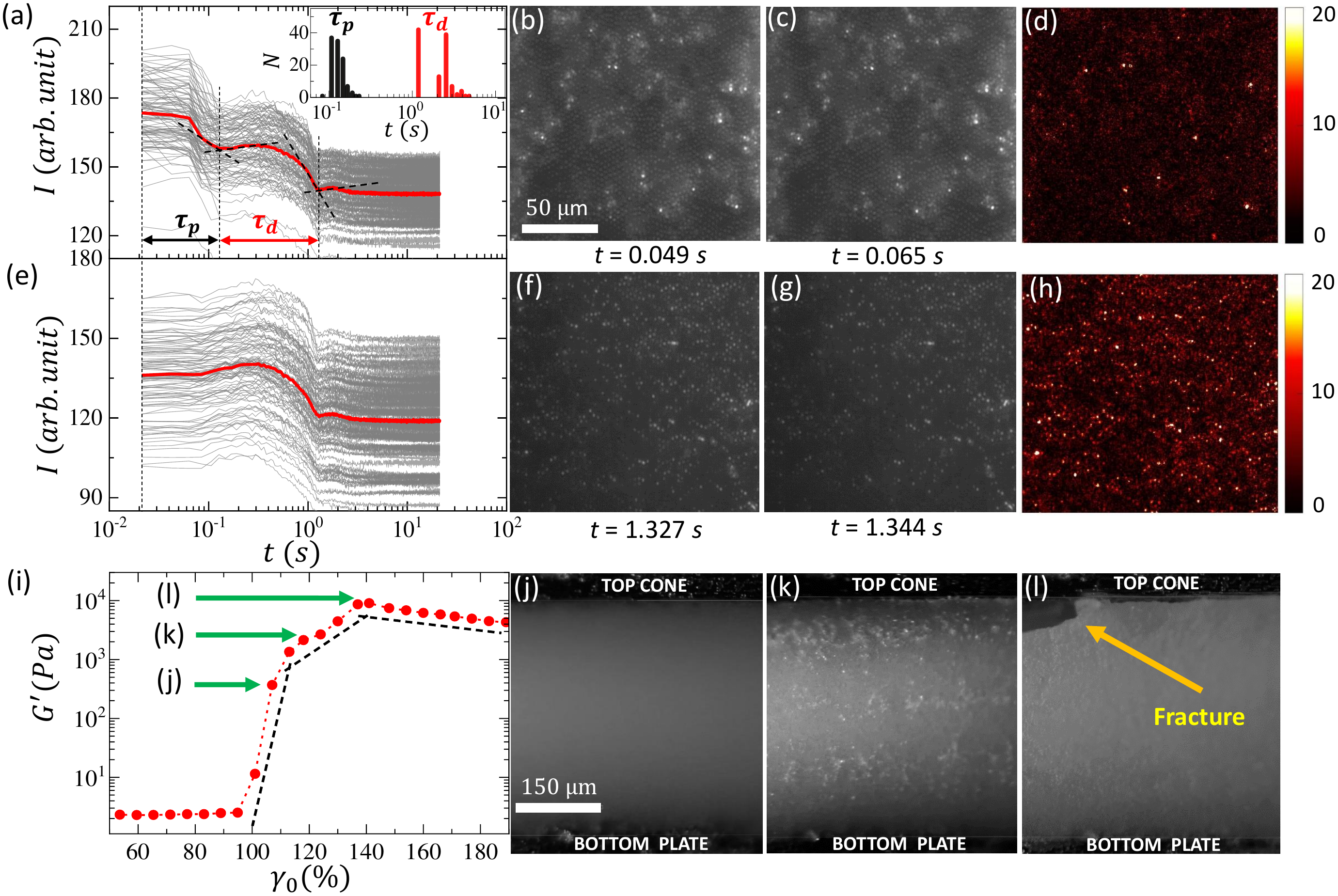}
\caption{\label{fig:wide} Intensity $I$ as a function of time $t$ (gray lines) for different regions with (panel (a)) and without (panel(e)) PCs with the average behaviors (thick red lines) are also indicated. Here, the starting time for each graph corresponds to an absolute time $t = t_0$ as mentioned in Fig.1. Relaxation time for PC ($\tau_p$) and dilation ($\tau_d$) are marked with vertical dash lines in (a). Inset: distribution of $\tau_p$ and $\tau_d$ ($N$ = 130). (b) and (c) Consecutive images of the sample boundary, during the fast relaxation and the corresponding difference image is shown in (d). (f) and (g) Consecutive images of the sample boundary, during slow relaxation and the corresponding difference image shown in (h). (i) Elastic modulus ($G'$) vs. strain amplitude ($\gamma_0$). Dashed lines represent the slope in different $\gamma_0$ regimes with the corresponding boundary images shown in panels (j), (k) and (l). $\phi=61$\%  and $\eta_l$ = 80 $mPa.s$ in all cases.}
\end{figure*}

To correlate the complex relaxation process with the particle scale dynamics, we use high-resolution in-situ optical imaging of the sample boundary in the flow-gradient plane \cite{bakshi2021strain}. For sufficiently high $\sigma_p$, we see an enhanced brightness of the sample boundary across the entire shear-gap (Movie S1) due to dilation \cite{brown2012role}. The boundary intensity returns back to the initial unperturbed value once the shear stress relaxes below $\sigma_d$. Surprisingly, for high $\sigma_p$ values, where the stress relaxation becomes discontinuous (beyond a critical magnitude of $\gamma$), we find randomly distributed bright spots (width: 1 - 2 particle diameter) [Fig.2(b) and 2(c) also, Movie S1 and S2] on the sample boundary. At high enough $\sigma_p$ these bright spots can combine to  create a macroscopic fracture in the sample (Fig.S5). Using oscillatory rheology, we find that the appearance of these bright spots is correlated with the onset of weakening/plasticity of SJ state, as marked by the decrease in slope of $G'$ vs. $\gamma_0$ (Fig. 2(i) - 2(l) and Fig.S6). This implies that these spots indicate localized plasticity/micro-fracture over the particle length scales and we term them as {\em{plastic centers}} (PCs). We find that the number of such PCs increases with $\gamma$ before fracture appears (Fig.S7). During stress relaxation, we find that PCs disappear quickly before a gradual decrease of surface intensity takes place. To quantify the time dependence of intensity relaxation we plot the variation of average intensity $I(t)$ as a function of time over a small region ($\sim 13.6\,\mu m^{2}$) around each PC (Fig.2(a)), as well as, similar regions that do not contain any PC (Fig.2(e)). For both of these regions, the intensity relaxation remains self-similar as indicated by the mean curves in Fig.2(a) and 2(e). For PCs we observe that $I(t)$ shows a two step relaxation similar to $\sigma(t)$. Due to small dynamic range of intensity relaxation, we define the associated timescales from the cross-over points (Fig.2(a)). We denote the fast time-scale as $\tau_p$ and the slower one as $\tau_d$. For regions without PCs, we find that the short-time drop in intensity is missing, but the slower relaxation behaviour is very similar to that obtained for PCs. Interestingly, we find that the values of $\tau_p$ and $\tau_d$ are distributed with peaks around $t \approx$ 0.2 s and 2 s, respectively (inset of Fig.2(a)) showing a strong  correlation with the stress relaxation timescales $\tau_f$ and $\tau_s$. This implies that the rapid stress relaxation of the SJ state is related to the dynamics of PC, whereas, the slower relaxation is governed by the dilation dynamics. Nonetheless, dilation is also present during the PC relaxation due to high stress in the system. During the PC relaxation, there is stress injection in the system due to fluidization of local jammed regions. This is reflected in sudden rearrangement and slight enhancement of the surface intensity indicating a stronger dilation beyond $t = \tau_p$ (Fig.2(e) and also Movie S2). This results in the peak in $N_1$ (around $t = 1$ s in Fig.1(c)). Such dynamic local jammed regions have also been observed in recent simulations \cite{seto2019shear}. To get a deeper insight into this striking intensity relaxation dynamics, we calculate the intensity difference between two consecutive images $\Delta I = |I(x,y,t) - I(x,y, t + \Delta t)|$ ($\Delta t$ = 0.016 s) during both PC (Fig.2(b) and 2(c)) and dilation (Fig.2(f) and 2(g)) relaxation. Appearance of a bright spot in $\Delta I$ at a particular spatial position indicates a local particle rearrangement at that position over a time $\Delta t$. We find that during PC relaxation $\Delta I$ shows only a few isolated bright spots (Fig.2(d)), whereas, during dilation relaxation we obtain a large number of bright spots uniformly distributed throughout the field of view (Fig.2(h)). This indicates that PC relaxation is governed by abrupt localized rearrangements, whereas, dilation relaxation happens by more gradual rearrangements throughout the system. We conjecture that since the PCs are sensitive to only local constrains, PC relaxation is  much faster compared to dilation relaxation involving global constrains. Such picture physically predicts the origin of the temporally distinct stress relaxation regimes. We do not observe any significant change in average particle distribution before and after the PC relaxation (Fig.S8), further confirming the spatially localized nature of PC relaxation. In our system, the interparticle contact formation time scale $\eta_l\,\gamma/\sigma* \sim 4 \times 10^{-4}$ s ($\sigma*$: onset stress for contact formation) is negligible compared to the relaxation time scales, implying that the relaxation is governed by the relaxation dynamics of contact networks \cite{richards2019competing, baumgarten2020modeling}. Importantly, such contact networks not only get stronger with increasing applied stress, but beyond a critical value significant stress induced reorganizations can happen through buckling and eventual breaking of the force chains \cite{cates1998jamming, baumgarten2019general} close to jamming. For rigid particle systems  such buckling can take place even at moderate stress values due to small area of contact (see S.I.). We also correlate the local plastic rearrangements with the discontinuous stress relaxation. We find an enhancement of the number of such rearrangements at the point of sharp stress drop (Fig.S9). Such correspondence has also been observed in the context of granular plasticity \cite{amon2012hot, le2014emergence}. In dry granular systems, X-ray tomography reveals that equivalent to nearest neighbour exchange (T1 events) in 2-D, particle-pair exchange neighbours resulting in defects of poly-tetrahedral order in 3-D \cite{cao2018structural}. However, we do not observe any T1 event from our 2-D imaging.

To test our conjecture regarding the fast and slow time-scales, we probe the effect of solvent viscosity ($\eta_l$) and particle volume fraction ($\phi$) on the stress/intensity relaxation time scales. We see from Fig.3(a) that the average value of fast time-scales obtained from the stress relaxation ($\tau_f$) and boundary imaging ($\tau_p$) remain almost independent of $\phi$. However, the slow time scales $\tau_s$ and $\tau_d$ show an increasing trend with the increase in $\phi$ before saturating for $\phi\geq$ 59\%. Similar trend is also observed for the change in solvent viscosity (Fig.3(a), inset): $\tau_f$ remains independent of $\eta_l$ but, $\tau_s$ increases with the increase in $\eta_l$.  These results imply that fast relaxation time-scales are governed by the local plasticity of the contact networks through the particle scale parameters, such as, surface roughness, rigidity, adhesion. On the other hand, slow time scales involving large scale rearrangements in the system should increase due to increase in drag (due to increase in $\eta_l$) or increase in the average coordination number (due to increase in $\phi$). Such behavior of longer timescale is also observed in simulation \cite{baumgarten2020modeling}. We find that the time scale obtained from the inverse of onset shear rate for shear-thickening under steady shear ($\tau_i = 1/\dot{\gamma_c}$) shows a good agreement with the slow timescales (Fig.3(a)), similar to flow cessation experiments \cite{maharjan2017giant}. These observations further confirm our conjecture about the origin of fast and slow time scales. 
\begin{figure}[h!]
\includegraphics[scale=.46]{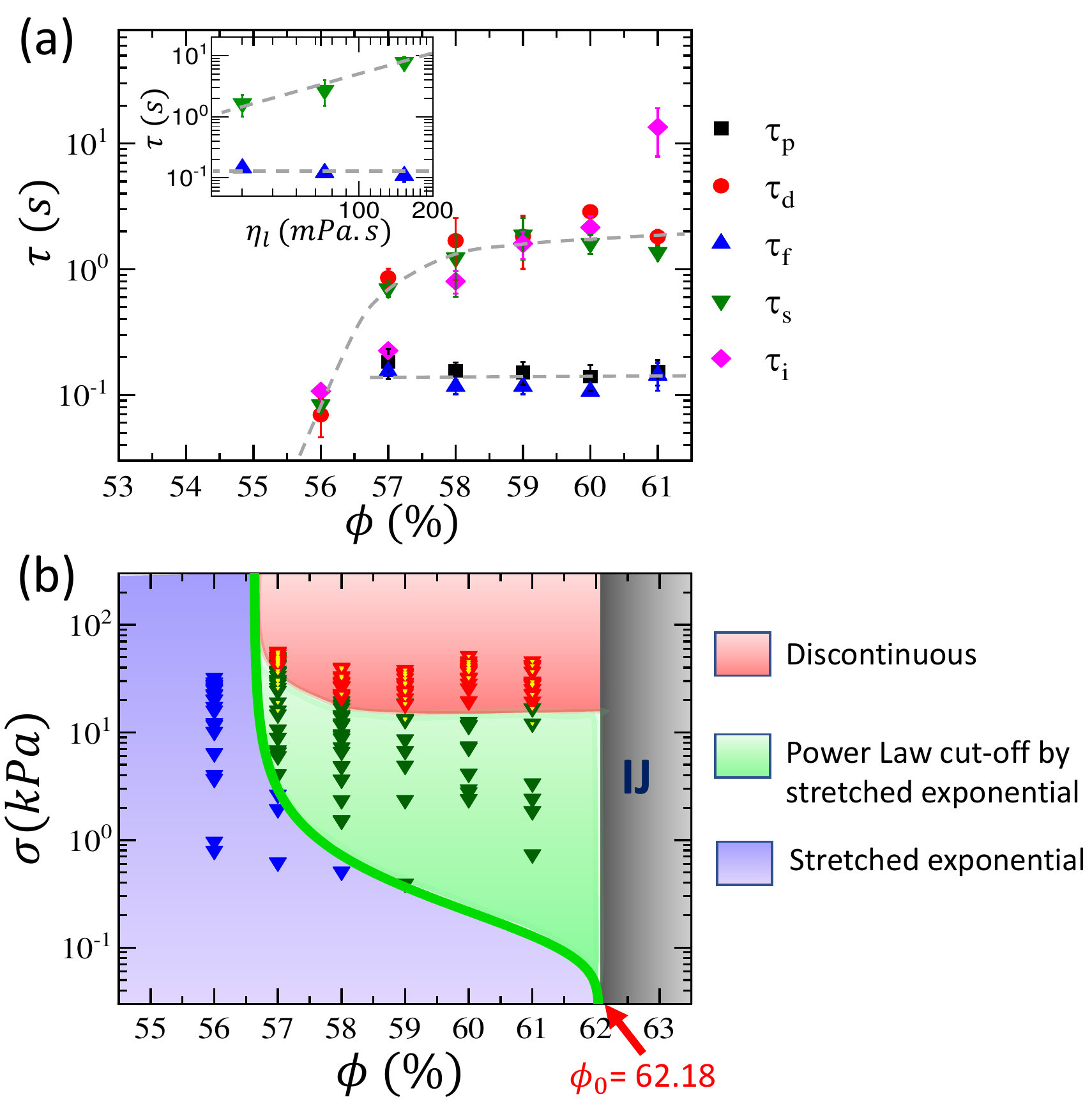}
\caption{\label{fig:epsart} (a) Dependance of fast ($\tau_p$ and $\tau_f$) and slow ($\tau_d$ and $\tau_s$) time scales as a function of volume fraction $\phi$. $\tau_i = 1/\dot{\gamma_c}$ obtained from the steady state measurements. Inset shows $\tau_f$ and $\tau_s$ as a function of $\eta_l$. Error bars indicate standard deviation over three independent measurements. Grey dashed lines are guides to the eye. (b) State diagram in $\sigma - \phi$ parameter plane. Thick green line indicates SSSJ onset. In the pink shaded region discontinuous stress relaxation (red triangles) and PCs (yellow dots) are observed, whereas, in green region (with dark-green triangles), a continuous relaxation (power-law cut-off by a stretched exponential) is found. Below SSSJ regime (purple region with blue triangles) initial power-law relaxation behavior disappears. Grey shaded region indicates isotropic jamming. In all cases symbols are the peak stress $\sigma_p$ obtained from the transient relaxation experiments.} 

\end{figure}
        
Finally, we construct a state diagram summarizing the results of transient stress relaxation in the $\sigma -\phi$ parameter plane (Fig.3(b)). The onset of SSSJ is obtained from the steady state flow curves (Fig.S1). As indicated in the diagram, sharp discontinuous stress relaxation (red region) over short time in the SJ regime is observed for high peak stress values $\sigma_p >$ 16 kPa. PCs also appear in this regime showing a strong correlation with the discontinuous stress relaxation. The capillary stresses at the solvent-air interface provide the confining stress over a wide range during shear-induced dilation in dense suspensions \cite{slobozhanin2006capillary, brown2012role, brown2014shear}. Since PC indicates abrupt local curvature due to significant protrusion of isolated particles/small clusters, the local confining stress at the PC can be approximated as the maximum capillary stress $\sim \Gamma/a \approx$ 16 kPa. Such capillary stress drives the protruding particle inside the bulk during the discontinuous stress relaxation over a time scale $\eta_t\, a/ \Gamma \sim$ 0.1 s ($\eta_t = \sigma_p/\dot{\gamma_p}$, see S.I. text and Fig.S10). This time scale is close to the fast relaxation time scale observed in our system. Intriguingly, in-between the discontinuous stress relaxation regime and onset of SSSJ (green region), we obtain a smooth relaxation regime (Fig.3(b)) where $\sigma(t) \sim t^{-\alpha}\, e^{-(t/\tau)^{\beta}}$. Below SSSJ, the initial power-law relaxation regime disappears.

In conclusion, we identify two distinct transient stress relaxation regimes in SJ dense suspensions originating from the dynamics of localized plasticity and system spanning dilation. Recently, an intrinsic contact-relaxation time-scales have been extracted from the coupling of relaxation with the instrument inertia \cite{richards2019competing}. Also, considering plasticity in the system, recent theoretical models \cite{baumgarten2020modeling} capture the short-time stress relaxation behavior. Although, the relaxation time scales in our system are within the range predicted in \cite{richards2019competing}, the robust initial power-law decay, presumably coming from stress induced force chain buckling/breaking can not be predicted from these models. We find that the fast and slow relaxation time scales are almost independent of step-strain magnitude [Fig.S11] and the estimated diffusion time-scales are inadequate to quantitatively capture them [S.I.]. Our preliminary data for larger polystyrene particles (mean diameter: 6.5 $\mu$m) also show similar relaxation dynamics. However, these directions including a possible extension into Brownian regime, require further exploration using more sophisticated experimental techniques \cite{pradeep2021jamming}. We find an interesting correlation between the transient stress relaxation and the steady-state shear jamming. The continuous stress relaxation showing a power-law cut-off by stretched-exponential behaviour is reminiscent of relaxation in frictionless systems close to jamming implying that for well-constraint systems the relaxation dynamics is not sensitive to the exact origin of the constrains. Such functional form indicates a wide range of underlying relaxation modes in the system \cite{hexner2018role}. We also observe similar relaxation behaviour including the discontinuous stress relaxation for other SJ dense suspensions, indicating an universal behaviour. Although, our study underscores the importance of local-plasticity in controlling the mechanical behaviour of SJ systems, deciphering the microscopic nature and dynamics of such plasticity, together with a possible connection to the more general framework of soft glassy rheology \cite{sollich1997rheology, le2014emergence, mohan2015build, chen2020microscopic} potentially unifying the relaxation behaviour in SJ and glassy systems, remains an important future direction.  

SM thanks SERB (under DST, Govt. of India) for support through a Ramanujan Fellowship. We thank Sidney Nagel for helpful discussions and K. M. Yatheendran for help with SEM imaging.

\clearpage

\begin{center}
{\Large{\textbf{Supplementary Information}}}
\end{center}
\section{\label{sec:level1} Sample Preparation and Experimental Protocol:}
 The shear thickening dense suspension is prepared by dispersing mono-disperse polystyrene (PS) particles (diameter: $2.65\pm 0.13$ $\mu m$) in a Newtonian solvent polyethylene glycol (PEG) over a wide range of volume fractions $40\% \leq \phi \leq61\%$. We use both PEG 400 and PEG 200 to vary the viscosity of the solvent. The viscosity can also be varied significantly by changing the temperature. We observe that above $\phi=61\%$ the particles are difficult to disperse in the solvent due to almost solid-like nature of the suspension. 
 
 The polystyrene micro-spheres are synthesized using dispersion polymerization in ethanol with styrene (TCI, Japan) as monomer, polyvinylpyrrolidone (PVP K-30, Spectrochem, India) as stabiliser and AIBN (Spectrochem, India) as initiator \cite{wang2013synthesis}. 
 
The samples are prepared by adding the desired quantity of PS particles in the solvent gradually and at each step the suspensions are mixed thoroughly using spatula. Next, the dense suspensions thus obtained are desiccated overnight to remove air bubbles, if any. To ensure an uniform dispersion, the samples are ultrasonicated for 5 minutes (at room temperature) just before loading for rheology measurements. 
  
All the rheological measurements are carried out on a stress control Rheometer (MCR-702, Anton Paar, Austria) with 12.5 mm radius cone and plate geometry. The surface of both the geometries are sandblasted to minimize the wall-slippage. The steady state measurements are performed in single drive and the transient measurements are done in twin drive under 50-50 counter movement mode. In order to remove the loading history and ensure that the sample condition remains unaltered after the experiment, we conduct large amplitude oscillatory shear (LAOS) measurement at a frequency $1$ Hz, before and after the stress relaxation experiment. For LAOS measurement we first increase the strain amplitude ($\gamma$) logarithmically from a lower to a maximum value and then we perform a reverse run where the amplitude is again gradually decreased to the initial value. We observe the variation of storage ($G'$) and loss ($G''$) moduli. We observe that the value of $G'$ and $G''$ as a function of $\gamma$ do not change significantly before and after the stress relaxation experiment. This also gives an idea of shear strain amplitude required to observe shear thickening in the sample.  
 For stress relaxation experiment under a transient perturbation, the sample is subjected to a step shear strain of a certain magnitude for 25 s. We record the resulting stress response of the sample over the same period of time. We increase the amplitude of the applied shear strain systematically, from lower to higher value, to drive the system into higher peak-stress levels gradually.

Since the particles are optically opaque, the in-situ imaging of the sample boundary is possible only under reflection mode. We use a Lumenera Lt545R camera with a 5X and 20X long working distance objective (Mitutoyo). During rheological measurements, the sample boundary is illuminated using a LED light source (Dolan-Jenner Industries) and we image the diffused scattering in the flow gradient plane \cite{bakshi2021strain} with frame rates varying between 60 Hz - 70 Hz.

\section{\label{sec:level1} Movie Descriptions:}
\subsection{\label{sec:level1} Movie S1:}
This video depicts the formation of plastic center (PC) along with the enhancement of the sample surface intensity due to dilation under transient step strain perturbation and the corresponding intensity relaxation obtained during the stress relaxation process. Here, a 70\% step strain is applied under counter-movement mode, where both the plates move equally in opposite directions. The images are captured using Lumenera Lt545R camera with a 5X long working distance objective (Mitutoyo) at a frame rate of 60 Hz, with resolution $1000\times1200$ pixels, in flow gradient plane. Here we notice that before the application of shear strain, the sample surface looks liquid-like, but just after the step strain application the overall intensity of the sample boundary surface increases due to dilation. Additionally, we find many tiny bright spots distributed throughout the surface. These tiny bright spots are called plastic centers (PCs) as described in the main text. During the stress relaxation process, we notice that these PCs are relaxing within a very short time scale, whereas, the overall intensity of the sample surface decreases to the initial value with a comparatively longer time scale. Here, the sample volume fraction $\phi = 61\%$ and the movie is played at 10 frames per second.

\subsection{\label{sec:level1} Movie S2:}
This video shows the formation of the PC along with dilation under step-strain perturbation and the corresponding intensity relaxation during stress relaxation process using a higher magnification. Here, we use a 20X long working distance objective (Mitutoyo) coupled to the camera. The images are captured at a frequency 60 Hz with a resolution of $800\times1000$ pixels. Higher magnification allows us to observe the individual particles undergoing dilation/forming PCs. We can clearly see that particle protrusions are more in PC regions as compared to dilation regions. In this case also the movie is played at 10 frames per second.

 \section{\label{sec:level2} Shear Jamming Boundary from Wyart-Cates Model:}
 The viscosity ($\eta$) of dense suspension can be expressed by Krieger-Dougherty (KD) equation as\\
\begin{align}
    \eta = \eta_l\left(1-\frac{\phi}{\phi_J}\right)^{-\beta}
\end{align}
Where $\phi_J$ is the jamming volume fraction, $\phi$ is the volume fraction of the sample, $\eta_l$ is the solvent viscosity and $\beta \approx 2$ for spherical particles \cite{singh2018constitutive}.

From Wyart and Cates model (WC model) the jamming volume fraction ($\phi_J$) can be written as
\begin{align}
    \phi_J = f(\sigma)\phi_m + [1-f(\sigma)]\phi_0
\end{align}
Where $f(\sigma) = e^{-(\sigma^*/\sigma)}$ denotes the fraction of frictional contacts, $\sigma^*$ is the onset stress for the frictional interaction, $\phi_0$ and $\phi_m$ represent the jamming volume fraction without any frictional contacts ($f=0$) and when all the contacts are frictional ($f=1$) respectively. From KD equation and WC model we find that $\phi_0 = 62.18\%$ and $\phi_m = 56.6\%$ as shown in Fig.S1(c) and S1(d) respectively.

Now, using $\eta = \sigma/\dot{\gamma}$ in Eq. 1, we can write
\begin{align}
    \dot{\gamma} = \frac{\sigma}{\eta_l}\left(1-\frac{\phi}{\phi_J}\right)^{\beta}
\end{align}

Considering the non-trivial solution ($\sigma \neq 0$) for $\dot{\gamma} = 0$ as the condition for shear jammed state and using Eq. 2, we can get
\begin{align}
    \phi = f(\sigma)\phi_m + [1-f(\sigma)]\phi_0
\end{align}
so,\begin{align}
f(\sigma) = \dfrac{\phi_0-\phi}{\phi_0-\phi_m}
\end{align}
Using  $f(\sigma) = e^{-(\sigma^*/\sigma)}$, we get
\begin{align}
   \boxed{\sigma = \frac{\sigma^*}{ln\left(\frac{\phi_0-\phi_m}{\phi_0-\phi}\right)}}
\end{align}
So, using Eq. 6 we can find the required stress ($\sigma$) for different volume fraction ($\phi$) to enter into the shear jammed state.

\section{\label{sec:level3} Estimation of area of contact between the particles:}
For a system of particles with particle stiffness $G_p$ the critical buckling load $\tau_c$ is given by \cite{baumgarten2019general} \\
\begin{align}
   \tau_c \sim G_p.\frac{A_c^2}{d^4} 
\end{align}
Where $A_c$ is the area of contact and $d$ is the particle diameter.

Now assuming the critical stress for chain buckling to be equal to the onset for discontinuous stress relaxation in our case, we have $\tau_c \sim 16000 Pa$ and diameter $d = 2.65 \mu m$.
For polystyrene particles considering Young’s modulus $\sim 3 \, GPa$ \cite{ref} we get $A_c \sim 16.2\times10^{-15}m^2$.
Again,
\begin{align}
  \frac{A_c^2}{d^4} = \frac{ \tau_c}{G_p.}
\end{align}
\begin{align}
\boxed{\frac{A_c}{d^2} = 2.3\times10^{-3}}\notag
\end{align}

Thus, we find that the area of contact is negligible compared to the surface area of the particles, as expected for rigid particles.
 
\section{\label{sec:level4} Estimation of plastic center relaxation time:}
From the imaging, we can clearly notice a significant protrusion of the particles at the PC regions in comparison to the surrounding. This indicates that the local stress ($\sigma_L$) at the PC approaches the maximum confining stress due to the solvent-air surface tension: $\Gamma /a$ ($a$: particle diameter , $\Gamma$: surface tension) \cite{brown2012role}.

When the particle is going back into the bulk it gives rise to an average local shear rate ($\langle \dot{\gamma_L} \rangle$). Thus, the local stress around the particle (assuming overdamped dynamics) is given by,
\begin{align}
\sigma_L=\eta_t \langle\dot{\gamma_L}\rangle  
\end{align}
Since PCs appear around the peak-stress ($\sigma_p$), we consider the transient viscosity ($\eta_t$) estimated from the ratio of peak stress ($\sigma_p$) developed in the system and the maximum shear rate ($\dot{\gamma_p}$) associated with the applied step strain.\\\\

During stress relaxation, the local stress around the particle should comparable to confining stress. Assuming the dynamics to be overdamped we can write:\\
\begin{align}
\sigma_L=\eta_t \langle\dot{\gamma_L}\rangle \sim \frac{\Gamma}{a}\notag\\
\dfrac{1}{\langle\dot{\gamma_L}\rangle} \sim \dfrac{\eta_t}{\left(\frac{\Gamma}{a}\right)}\notag\\
\boxed{t \sim \dfrac{\eta_ta}{\Gamma}}
\end{align}
For our system, $\Gamma \approx 44$ $mN/m$, $a=2.65$ $\mu m$. The time scales estimated from Eq. 10 is shown in Fig.S10a for different $\phi$ values. Also, as shown in Fig.S10b, $\eta_t$ depends very weakly on solvent viscosity.\\ For comparison we also estimate $\tau_p$ using maximum  viscosity ($\eta_{max}$) obtained from steady state measurements (Fig.S10a).\\ When we replace $\eta_t$ in Eq. 10 with the solvent viscosity $\eta_l$, we get $\tau_p\sim 5\times 10^{-6}$ s which is far below the experimental observation. \\We want to point out that the fast relaxation time scales observed in our case also have some contributions from stress induced reorganization in the system close to jamming (presumably through buckling/breaking of force chains). However, in the estimate mentioned above, we ignore such contributions which are currently unknown in our case.

\section{\label{sec:level5} Diffusion time scales for the system}
We show the variation of the fast and slow relaxation time scales for a range of particle volume fraction and applied step-strain magnitude in Fig. S11a. We find that both the time-scales remain almost independent of applied step strain magnitude.

Since our system is weakly Brownian (mean particle diameter $2r$ = 2.65 $\mu$m), the intrinsic diffusion coefficient ($D$) at room temperature is very small: $D = \frac{k_B T}{6 \pi \eta r} \sim 10^{-15} m^2/s$. Here we have assumed the value of $\eta$ to be the solvent $\sim$ 0.1 Pa.s. If we assume the $\eta$ to be the transient viscosity of the suspensions (Fig. S10), $D \sim 10^{-18} m^2/s$. Thus, intrinsic diffusion time-scale (in 3-D) $t_D \sim \frac{r^2}{6D}$ is orders of magnitude greater than the slower time scale of the system. For soft glassy systems, it is observed that the particle diffusion coefficient under shear has a dependence on associated strain rate and particle diameter ($d$) as $D\sim d^2 \dot{\gamma}$ \cite{utter2004self}. Other studies have also found similar relation $D\sim \dot{\gamma}^{0.8}$ \cite{besseling2007three} for higher shear rate and  $D\sim \dot{\gamma}^{0.78}$ at the jamming transition \cite{olsson2010diffusion}. However, for our case, as the stress relaxation takes place at constant strain, the strain rate is zero during stress relaxation process.

Using a similar formulation as that mentioned in \cite{eisenmann2010shear}, we have also checked the particle diffusivity in our system by replacing the thermal energy with the shear energy. Replacing the thermal energy term $k_B T$ with the shear energy $\frac{4\pi r^3\eta \dot{\gamma}}{3}$ with the assumption that the characteristic volume is equal to that of a particle, we can write the diffusivity as,
$D = \frac{\frac{4\pi r^3\eta \dot{\gamma}}{3}}{6\pi \eta r} = \frac{2 r^2 \dot{\gamma}}{9}$.
                                                       
Now, the diffusion time scale (in 3-D) will be  $t_D = r^2/6D = \frac{3}{4 \dot{\gamma}}$ 

From this expression, it is clear that the diffusion time scale is only depends on strain rate. The suspension viscosity and particle diameter play no role.

Although, in our case the strain rate remains zero during the relaxation process, we have considered the peak strain rate ($\dot{\gamma_p}$) associated with the applied step-strain. However, as we see from Fig.S11, both the fast and slow relaxation time scales are almost independent of the peak strain rate. Also, the values of the diffusion time-scales obtained for the range of peak strain rate values probed in our experiments ($\sim$ 10 to 50 $s^{-1}$, see Fig. S11) come out to be in the range 0.015 to 0.075 s which is significantly smaller than the time-scales observed in our experiments. This indicates that the colloidal time scales are not quantitatively capturing the observed relaxation time scales for the shear jammed state.

\bibliography{Supli}

\bibliographystyle{apsrev}

\newpage
\begin{figure*}[h]
\centering
\renewcommand{\figurename}{Fig.S1}
\renewcommand{\thefigure}{}
\includegraphics[scale=.55]{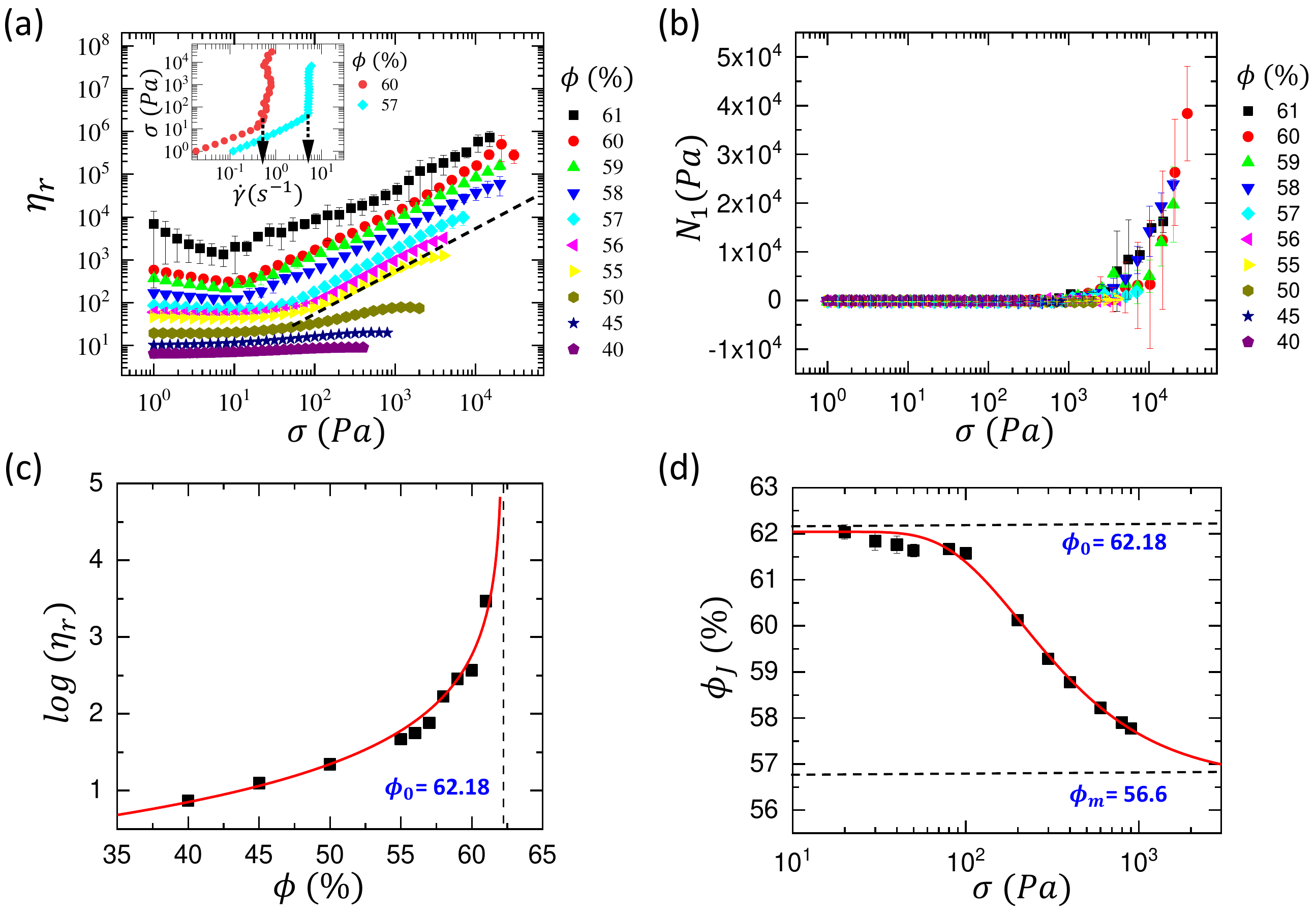}
\caption{\label{fig:wide} Results from the steady state measurements. (a) Variation of relative viscosity $\eta_r$ with shear stress $\sigma$ for volume fraction $40\%\leq\phi\leq61\%$ as shown in the legend. Black dash line represents a slope of 1. Error bars represent the standard deviation over three independent measurements. Inset shows shear stress $\sigma$ as a function of shear rate $\dot{\gamma}$ for volume fraction $60\%$ and $57\%$ as shown in the legend. The dotted arrows indicate the critical shear rate $\dot{\gamma_c}$ for the corresponding volume fractions. (b) Variation of the first normal stress difference $N_1$ with shear stress $\sigma$ during the steady state measurements shown in (a).  (c) Variation of $log(\eta_r)$ with $\phi$ (symbols) for $\sigma = 20$ $Pa$ as obtained from (a). Red solid line in (c) indicates the fit to the Krieger Dougherty (KD) relation (Eq. 1) with $\beta = 1.9$. Vertical dash line corresponds to jamming volume fraction without frictional contact $\phi_0 = 62.18\%$ determine from the KD equation fitting. (d) Variation of jamming volume fraction $\phi_J$ with $\sigma$ (symbols). Red solid line indicates the fit to the Wyart-Cates (WC) model (Eq. 2). Top and bottom horizontal dash line indicate the $\phi_0=62.18\%$ and $\phi_m =56.6\%$, obtained from the WC model fitting.}
\end{figure*}
\begin{figure*}
\centering
\renewcommand{\figurename}{Fig.S2}
\renewcommand{\thefigure}{}
\includegraphics[scale=.55]{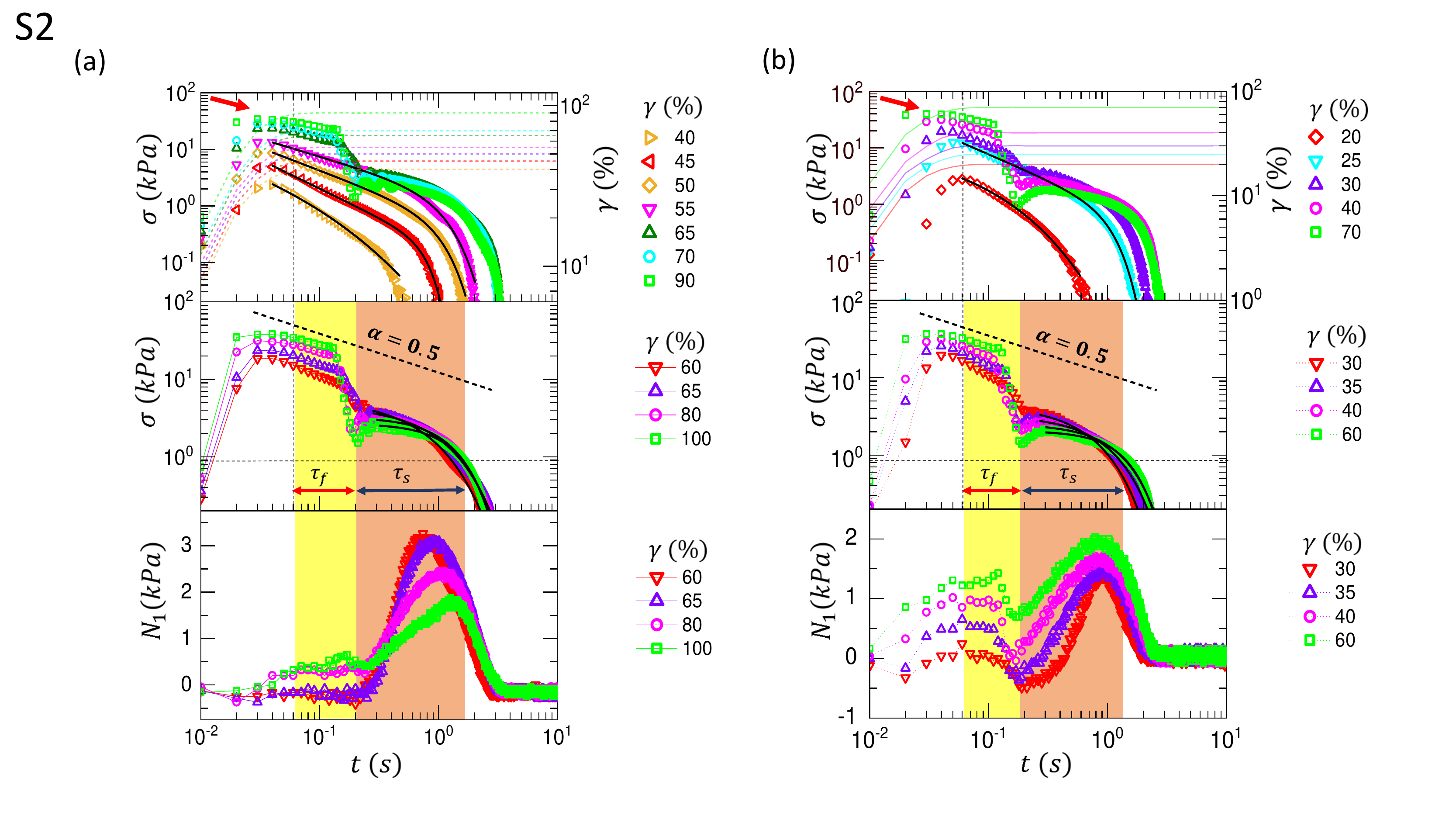}
\caption{\label{fig:wide} Variation of shear stress $\sigma$ and first normal stress difference $N_1$ (similar to that shown in Fig.1 in the main text) for volume fraction $\phi = 59\%$ (panel (a)) and $\phi = 60\%$ (panel (b)).}
\end{figure*}
\begin{figure*}
\centering
\renewcommand{\figurename}{Fig.S3}
\renewcommand{\thefigure}{}
\includegraphics[scale=.55]{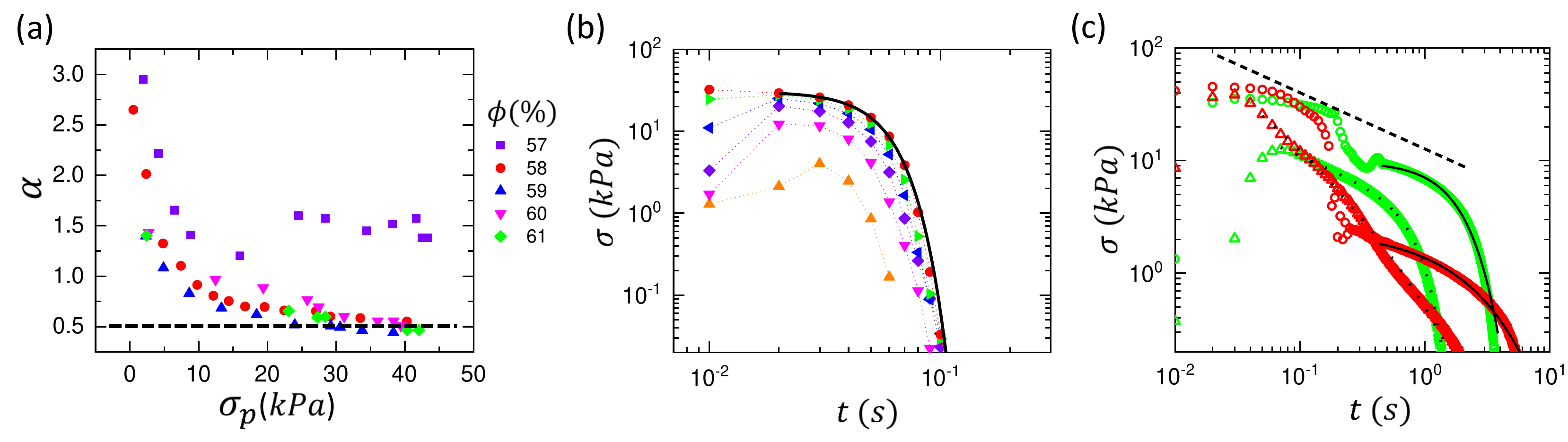}
\caption{\label{fig:wide} (a) Variation of power-law exponent $\alpha$ with peak stress $\sigma_p$ for different $\phi$ indicated in the legend. We find that $\alpha$ decreases from higher value with increase in $\sigma_p$ and saturates around 0.5 (except for $\phi=57\%$ showing saturation around $\alpha = 1.5$), as indicated by the horizontal dashed line.
(b) Shear stress $\sigma$ as a function of time $t$ (symbols with dotted line) for $\phi=56\%$. Here the initial power-law behaviour disappears and a stretched exponential  function $(exp(-(\frac{t}{\tau})^{\beta}))$ fits the data well (solid line).  (c) $\sigma$ as a function of $t$ (symbols) for two well known shear jamming system, the dense suspension of corn starch in water for $\phi = 50.1\%$ with swelling corrections (green) and silica sphere in water glycerol mixture for $\phi = 55\%$ (red). Both the systems show the discontinuity during stress relaxation for sufficiently high value of $\sigma_p$ where second part of relaxation can be fitted well with stretched exponential function (solid line). On the other hand, for the continuous stress relaxation, a power-law cut-off by a stretched exponential function fits reasonably well (dotted line). Dashed line corresponds to $\alpha = 0.5$.}
\end{figure*}
\begin{figure*}
\centering
\renewcommand{\figurename}{Fig.S4}
\renewcommand{\thefigure}{}
\includegraphics[scale=.65]{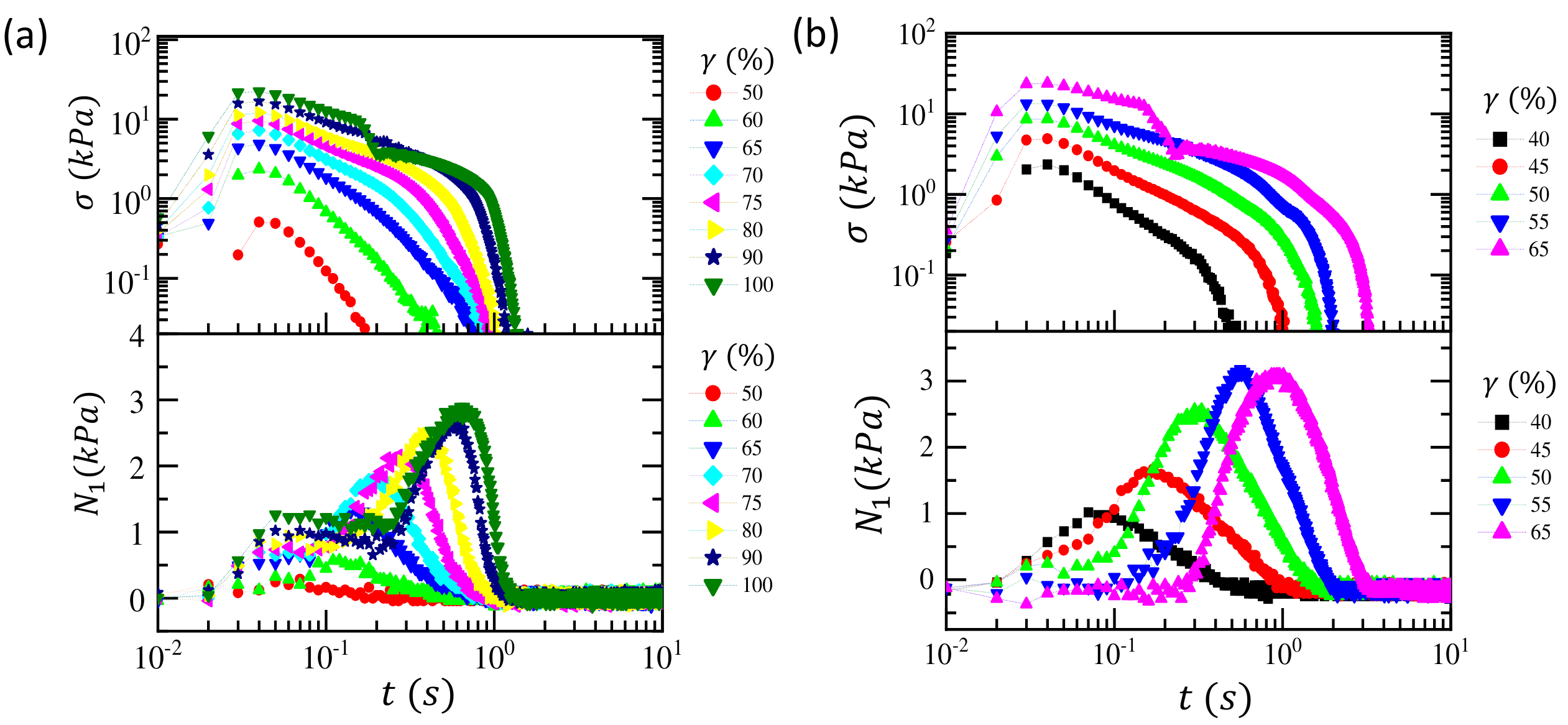}
\caption{\label{fig:wide} Variation of first normal stress difference $N_1$ with time $t$ for applied strain values $\gamma$ where continuous stress relaxation is observed. Here, volume fraction $\phi = 58\%$ (panel(a)) and $\phi = 59\%$ (panel(b)). For comparison only one discontinuous relaxation data is shown for each volume fraction. We see that the magnitude and peak position of $N_1$ are controlled by the magnitude of the peak stress in the system.}
\end{figure*}
\begin{figure*}
\centering
\renewcommand{\figurename}{Fig.S5}
\renewcommand{\thefigure}{}
\includegraphics[scale=.5]{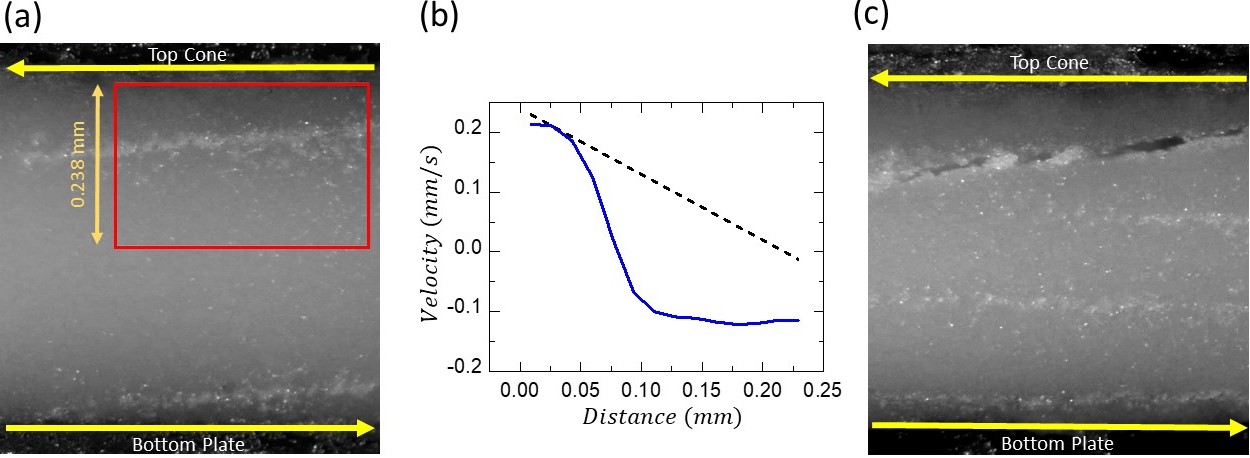}
\caption{\label{fig:wide} (a) Plastic center accumulation along a line (shear strain $\gamma = 70\%$). Such line of PC under high stress forms a macroscopic crack.  Red box represents the region of interest (ROI) for the velocity profile calculation around the line of PC with the velocity profile shown in (b). In panel (b), blue solid line represents measured velocity profile across the ROI which significantly deviates from the linear velocity profile (black dashed line), indicating strong non-affine deformation across the line of plastic centers. (c) Image of macroscopic fracture in the sample under large step strain deformation (shear strain $\gamma = 120\%$).}
\end{figure*}
\begin{figure*}
\centering
\renewcommand{\figurename}{Fig.S6}
\renewcommand{\thefigure}{}
\includegraphics[scale=.5]{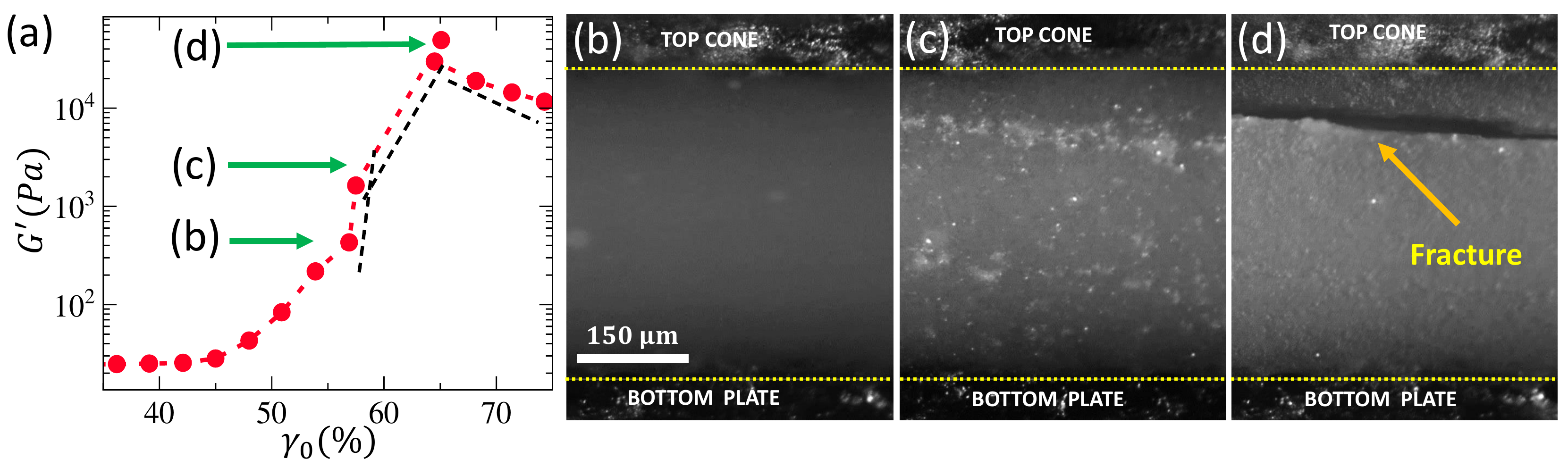}
\caption{\label{fig:wide} (a) Variation of elastic modulus ($G'$) with strain amplitude ($\gamma_0$). The dashed lines represent the slope of the curves in different $\gamma_0$ regimes. The surface images on the right panels correspond to three different points marked as (b),(c) and (d) (also indicated by arrows) in (a). Image around point (b) shows no signature of plasticity whereas significant bright spots start appearing around point (c) with the decrease in slope of $G'$ as a function of $\gamma_0$. These bright spots combined to form macroscopic fractures that result in a decrease of $G'$ beyond point (d). The change in slope of $G'$  around point (c) represents the weakening of the system with the appearance of microscopic plasticity in the form of bright spots which leads to macroscopic fracture in the system.}
\end{figure*}
\begin{figure*}
\centering
\renewcommand{\figurename}{Fig.S7}
\renewcommand{\thefigure}{}
\includegraphics[scale=.8]{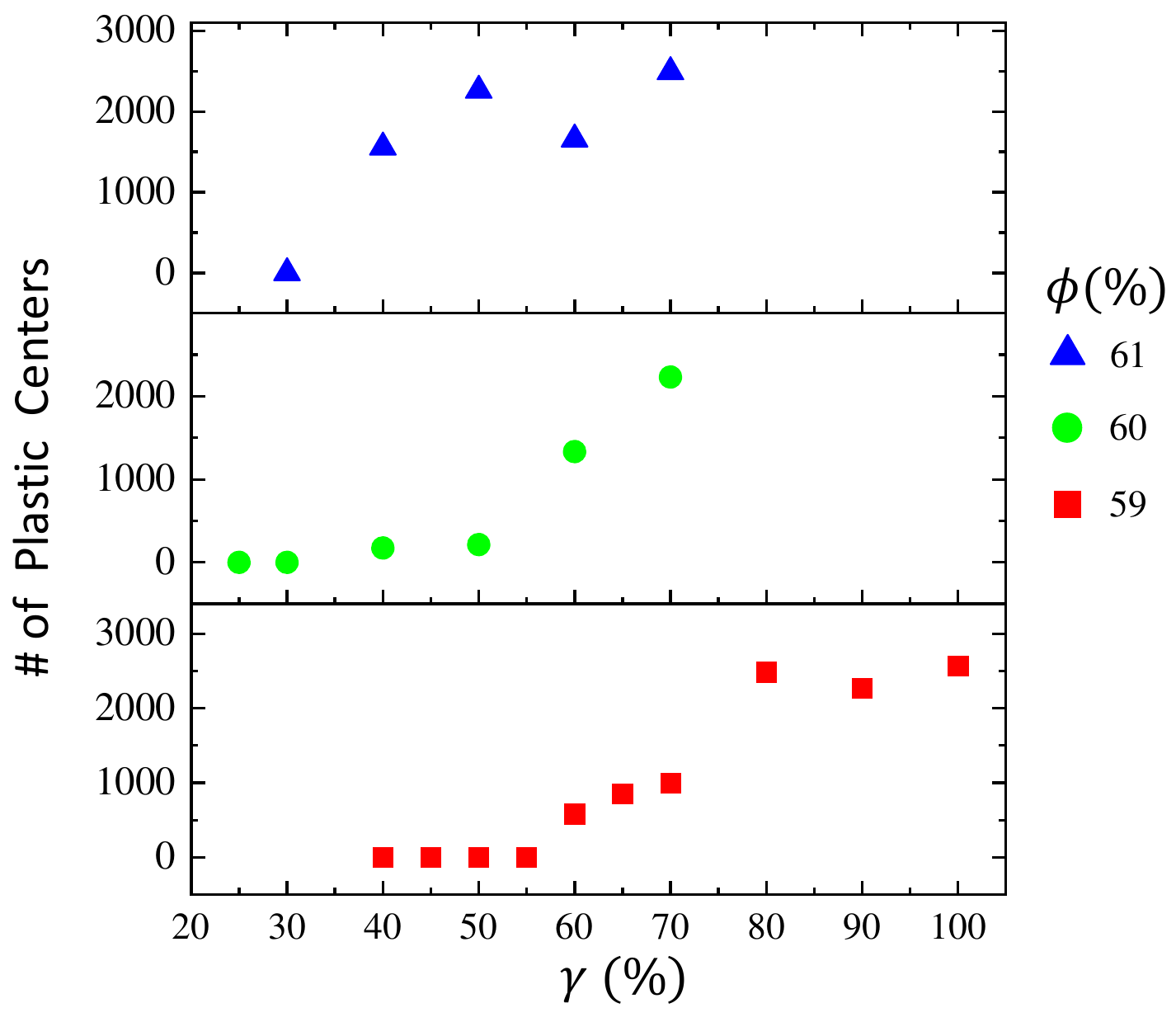}
\caption{\label{fig:wide} Number of plastic centers as a function of applied step strain $\gamma$ for different volume fraction $\phi$, as shown in the legend. For each $\phi$ value the number of plastic centers increases with increasing $\gamma$ beyond a strain onset.}
\end{figure*}
\begin{figure*}
\centering
\renewcommand{\figurename}{Fig.S8}
\renewcommand{\thefigure}{}
\includegraphics[scale=.50]{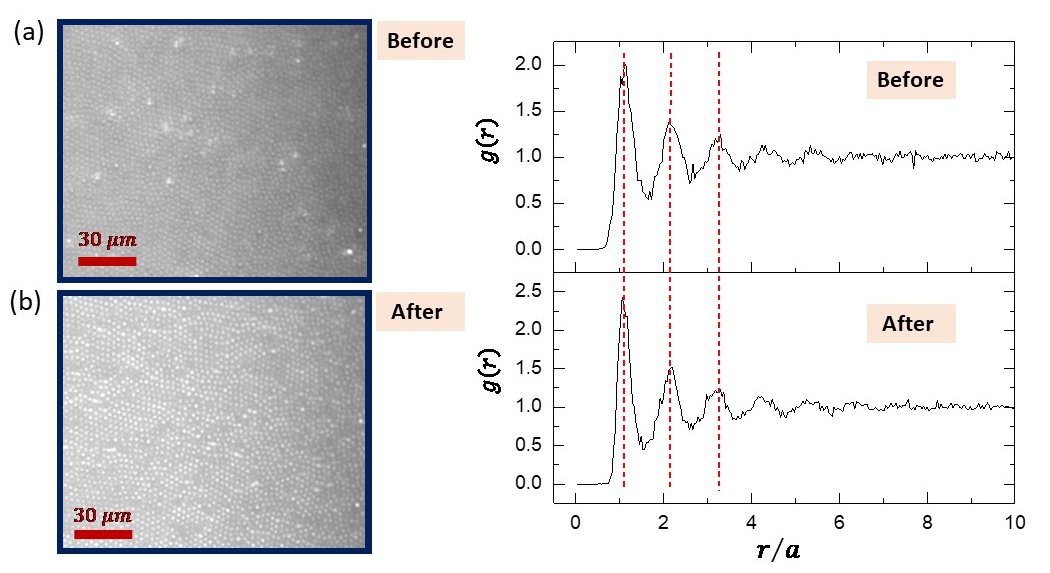}
\caption{\label{fig:wide} Boundary images before (panel (a)) and after (panel (b)) the PC relaxation. The corresponding radial distribution function $g(r)$ as a function of $r/a$ where $a$ is the diameter of the particle. Similar peak positions indicate that there is no significant difference in particle arrangements before and after the PC relaxation. In both the cases the $g(r)$ shows a short-range correlation that decays over a few particle diameters indicating the absence of any crystalline order in the system.}
\end{figure*}
\begin{figure*}[htp]
\centering
\renewcommand{\figurename}{Fig.S9}
\renewcommand{\thefigure}{}
\includegraphics[scale=.9]{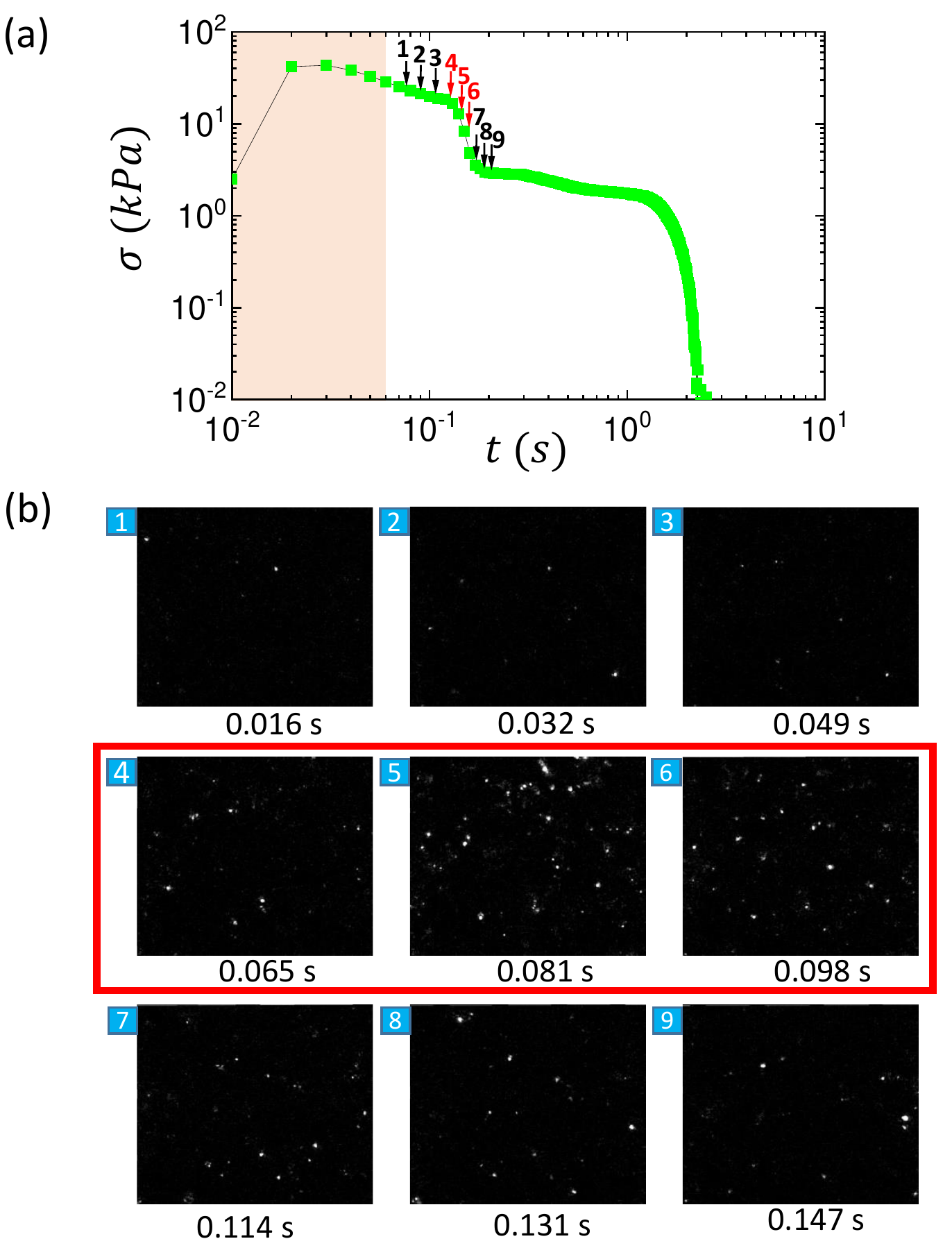}
\caption{\label{fig:wide} (a) Relaxation of shear stress $\sigma$ with time $t$ for a transient step strain experiment. (b) The stroboscopic difference images obtained from two consecutive images (time separation $\approx$ 16 ms) at different points during fast relaxation process (indicated by the different arrows in (a)). The increased number of isolated bright spots for image number 4, 5 $\&$ 6 (indicated by the red box) indicate an enhancement of plastic rearrangements near the sharp stress drop (indicated by the red arrows in (a)).}
\end{figure*}
\begin{figure*}[htp]
\centering
\renewcommand{\figurename}{Fig.S10}
\renewcommand{\thefigure}{}
\includegraphics[scale=.65]{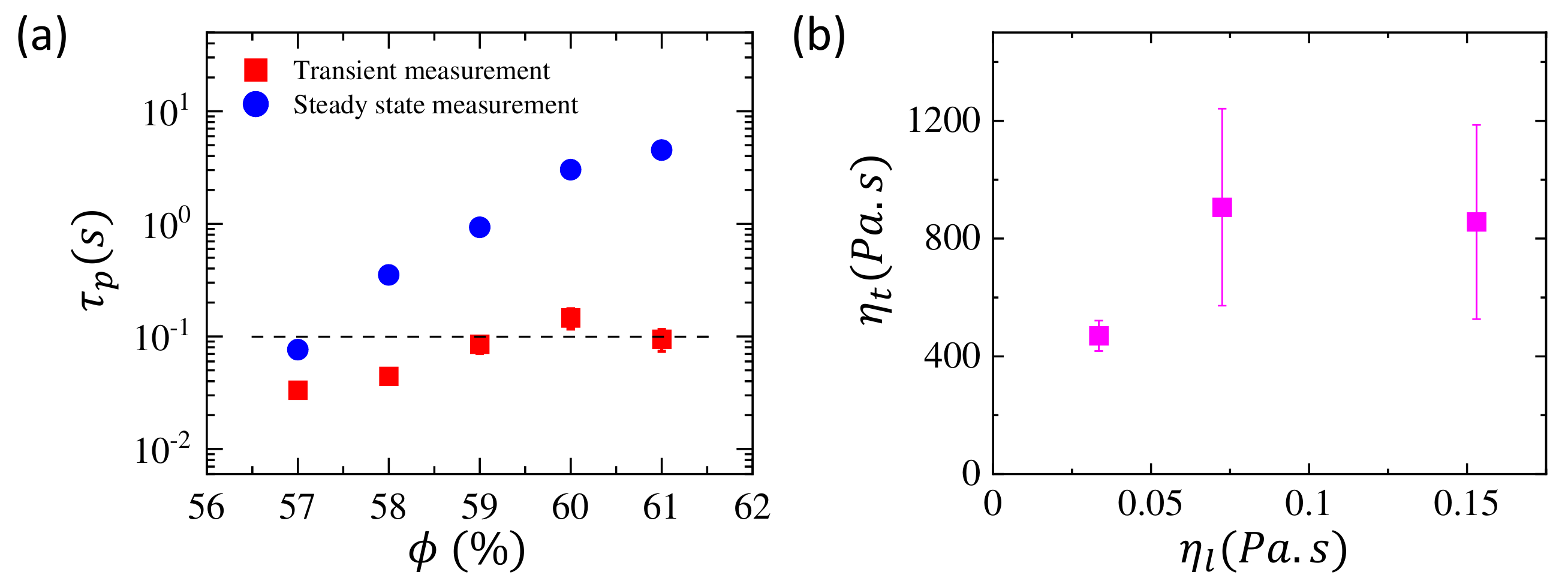}
\caption{\label{fig:wide} Variation of plastic center relaxation time $\tau_p$, calculated using Eq. 10, with volume fraction $\phi$. Red squares represent $\tau_p$ obtained using viscosity ($\eta_t$) from transient measurements. For comparison we also estimate $\tau_p$ using maximum  viscosity ($\eta_{max}$) from steady state measurements (blue circles). Horizontal dashed line indicates the timescale of 0.1 s. (b) Transient viscosity ($\eta_t$) obtained during the step stain measurements as a function of solvent viscosity $\eta_l$. We see that $\eta_t$ weakly depends on $\eta_l$.}
\end{figure*}
\begin{figure*}[htp]
\centering
\renewcommand{\figurename}{Fig.S11}
\renewcommand{\thefigure}{}
\includegraphics[scale=.5]{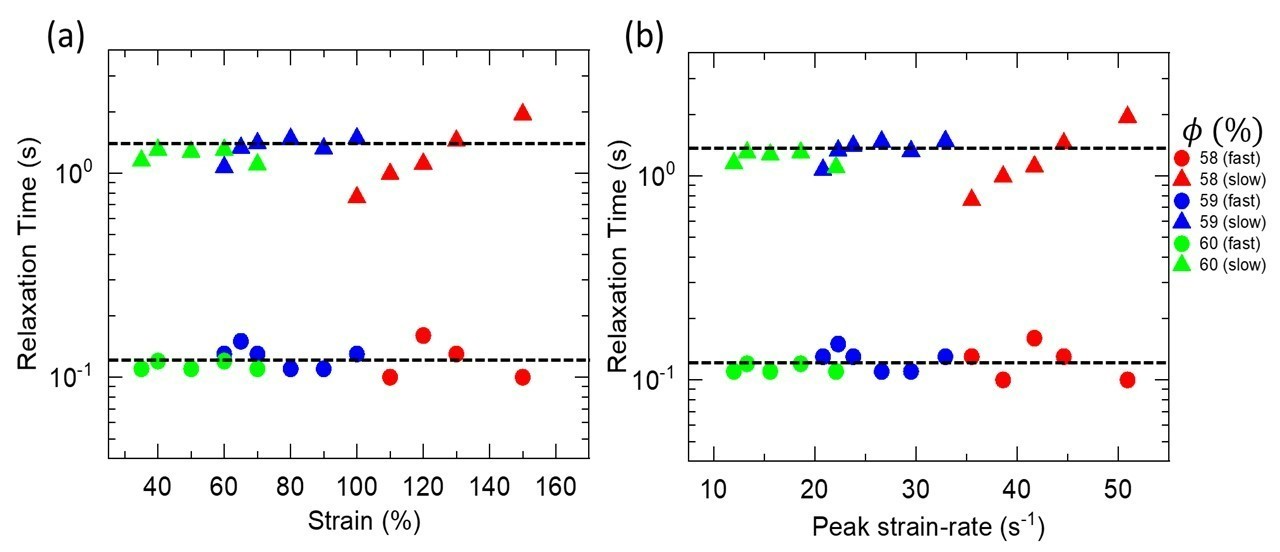}
\caption{\label{fig:wide} Variation of fast (circles) and slow(triangles) relaxation time scales with step strain magnitude (panel (a)) and the peak strain rate (panel (b)) for volume fractions as indicated. The horizontal dashed lines represent the fast and the slow relaxation time for $\phi$ = 59 \%. The relaxation time scales are almost independent of step strain magnitude, as well as, peak strain rate.}
\end{figure*}

\end{document}